# Pre-frontal cortex guides dimension-reducing transformations in the occipito-ventral pathway for categorization behaviors

Yaocong Duan,[1] Jiayu Zhan,[1] Joachim Gross,[2] Robin A.A. Ince,[1] & Philippe G. Schyns[1,3,4,*]

[1]School of Psychology and Neuroscience, University of Glasgow, 62 Hillhead Street, Glasgow, G12 8QB, United Kingdom

[2]Institute for Biomagnetism and Biosignalanalysis, University of Münster, Malmedyweg 15, Münster 48149, Germany

[3]X (formerly Twitter): @SchynsPhilippe

[4]Lead contact

*** Correspondence: philippe.schyns@glasgow.ac.uk (P.G.S.)**

Philippe G. Schyns
School of Psychology and Neuroscience
56, Hillhead Street
University of Glasgow
Glasgow G12 8QB
United Kingdom
+44 141 330 4937

**Summary**
To interpret our surroundings, the brain uses a visual categorization process. Current theories and models suggest that this process comprises a hierarchy of different computations that transforms complex, high-dimensional inputs into lower-dimensional representations (i.e. manifolds) in support of multiple categorization behaviors. Here, we tested this hypothesis by analyzing these transformations reflected in dynamic MEG source activity while individual participants actively categorized the same stimuli according to different tasks: *face expression, face gender*, *pedestrian gender*, *vehicle type*. Results reveal three transformation stages guided by pre-frontal cortex. At Stage 1 (high-dimensional, 50-120ms), occipital sources represent both task-relevant and task-irrelevant stimulus features; task-relevant features advance into higher ventral/dorsal regions whereas task-irrelevant features halt at the occipital-temporal junction. At Stage 2 (121-150ms), stimulus feature representations reduce to lower-dimensional manifolds, which then transform into the task-relevant features underlying categorization behavior over Stage 3 (161-350ms). Our findings shed light on how the brain's network mechanisms transform high-dimensional inputs into the specific feature manifolds that support multiple categorization behaviors.

## Introduction

Despite the intricate and detailed nature of visual input, our ability to categorize relies on extracting and processing the essential elements of this information–i.e. the features that are crucial for the task at hand. For example, whereas categorizing the scene in Figure 1A as a "happy face" requires processing the mouth of the central face, categorizing this same picture as a "SUV" requires processing the shape of the right-flanked vehicle, or the left "female pedestrian" with the bodily features that disclose its gender. The key point is that a single input image, and even a single object within this image, typically affords multiple different categorization behaviors (e.g. "happy," or "female" for the same central face), each relying on varying sets of features. Consequently, the internal representation of a single visual input can encompass multiple manifolds that vary depending on the task. Thus, when our brain categorizes the visual inputs, it doesn't passively represent the entire space of these inputs. Instead, current theories and models suggest that brain networks actively transform the representations of the same complex input images into specific low-dimensional manifolds (geometric subspaces of the images) that comprise diagnostic features for the task.[1–9] Here, we test this fundamental hypothesis, by reverse engineering, at a system level, the dynamic brain networks that actively represent and transform identical input scene images for distinct categorization behaviors.

Significant progress in understanding visual categorization resulted from accurately mapping the brain regions that respond to various categories of images[10–13] (e.g. those of faces, bodies, objects and scenes). These regions comprise primarily the occipito-ventral/dorsal pathways that respond to different image categories, from their early split projection in left and right occipital cortex to their later categorical/semantic representations in the right fusiform gyrus, including how feedback reverses this flow to predict the input stimulus.[14–17] Additionally, Pre-Frontal cortex (PFC) can represent categories separately, which minimizes their interference.[18] This separation allows for task focus and attention to modulate object selectivity,[19,20] enhance visual processing[21] and improve semantic representation, all of which optimize behavior.[22] Though this approach proved invaluable to investigate <u>where</u> and <u>when</u> different brain regions are involved with processing full images, it overlooked <u>how</u> the task itself changes the feature manifolds that the brain processes from input images for categorization behavior.

Research into eye movements,[23,24] attention,[25,26] reverse correlation[4,5,27] and neural network modelling[2] suggests that categorization mechanisms in capacity-limited systems actively and flexibly transform high-dimensional input images into the low-dimensional feature manifolds representing the visual information that support task-specific behavior (e.g. the smiling mouth feature for categorizing a *facial expression* as “happy”, or the sedan shape feature for categorizing *vehicle type* as “sedan” from the same image in Figure 1A), guided by frontal-parietal network mechanisms.[28] Critically, what emerges is an active process whereby brain networks process features that are not inherently given, but instead dynamically extracted from the image depending on the participant’s categorization task and their individual strategy.[29]

To track such image transformations into dynamic neural responses requires a broad, systems-level approach with fine-granularity control of the stimuli. Stimulating with categories of uncontrolled faces, vehicles and pedestrians full images, as is typical,[30–32] makes it practically unfeasible to precisely track what specific features the participant’s brain processes for behavior. Instead, these features are latent in the image pixels.[2,6] To reveal the categorization features, we applied the Bubbles procedure.[7,27] Bubbles randomly samples the pixels of a stimulus image with Gaussian apertures, that each participant then categorized in four different ways–i.e. as *face expression*, *face gender*, *pedestrian gender* and *vehicle type* (see Figure 1A). Bubbles ensures that the participant can only correctly categorize when the randomly sampled image pixels show the features needed for categorization.[7,27] With such control, we could reverse engineer (1) the feature manifolds that each participant processes for behavior in each task[33] and critically, (2) where (networks of brain regions), when (which time windows) and how (with what transformations) the activity of 5,107 cortical MEG sources (every 1.67 ms between 0 to 450 ms post stimulus) transformed their representation of the same images into task-specific feature manifolds that support behavior.

To preview our findings, we provide a detailed descriptive model of how the categorization task modulates internal transformations of the visual input over three systems-level Stages. At Stage 1 (high-dimensional, 50-120 ms), occipital sources represent more stimulus features than the task requires–i.e. including occipital source-level opponent representations[34–37] of a feature when it is task-relevant vs. irrelevant. While task-relevant features advance into ventral/dorsal pathways,[38–42] irrelevant ones are halted at the occipital-temporal junction. At Stage 2 (high-to-low dimensional, 121-150 ms), occipital sources reduce most irrelevant features, while ventral-dorsal pathways represent manifolds that keep transforming over Stage 3 (low-dimensional 161-350 ms) into the task-relevant features[28,43–46] underlying categorization behavior (e.g. smiling mouth features underlying “happy” vs. sedan features underlying “sedan”). Furthermore, Pre-Frontal Cortex (PFC) interacts with occipital-ventral/dorsal pathways early on (during Stages 1-2, 71-150 ms post-stimulus), suggesting network mechanisms involving these regions that transform stimulus image representations into task-relevant feature manifolds based on the task.

## Results

Our experiment comprised four 2-Alternative-Forced-Choice (AFC) categorization tasks applied to the same 64 base images of a realistic, complex city street scene (see Figure 1A). These images comprised varying embedded targets–i.e. 8 different face identities[47] (2 genders), each representing 2 expressions x 2 vehicles x 2 pedestrians. Each participant (N = 10, within-participant statistics[48]) performed each 2-AFC task in different blocks of 1,536 trials (i.e. precision neuroscience with dense sampling[49]). Figure 1A illustrates, with two examples of the base images, the combinatorics of stimulus and 2-AFC task-response differences (i.e. in *face expression*, *face gender*, *pedestrian gender* and *vehicle type*).

Each trial started with a fixation cross presented for a random time interval between 500-1000 ms, followed by one base image for 150 ms, whose pixels were randomly sampled with Bubbles[7,27] (see Figure 1 and *Methods, Stimuli*). Random Bubbles sampling ensures the participant can only correctly categorize the stimulus when the samples reveal by chance the pixels of the features the participant requires to resolve the task. For example, the randomly sampled pixels of trial *n* in Figure 1 would categorize “happy” in the *face expression* block, but not “SUV” in the *vehicle type* block, and vice versa with the samples of trial *m* (see *Methods, Task Procedure*). Critically, the set of randomly sampled stimuli was identical in each participant and blocked task (with stimuli presented in a random order in each block), eliminating all low-level artefacts when different stimuli are associated with different categorizations. On each trial, we concurrently recorded the participant’s dynamic MEG activity (localized with a beamformer to 5,107 cortical sources, see *Methods, MEG*) and categorization responses.

**Behavior: Task-relevant feature manifolds**

To reconstruct the categorization-feature manifolds supporting task performance (see Tables S1 and S2), in each participant we quantified the cross-trial relationship between pixel presence vs. absence due to random sampling (Figure 1C) and corresponding behavioral responses in each task—computed with Mutual Information (MI[50,51]) as MI(pixel visibility; correct vs. incorrect categorization), controlling the Family-Wise Error Rate (FWER) $p < 0.05$ over pixels, see *Methods, Analyses, Participant features*.

Figure 1C shows that participants use task-specific features from an identical set of sampled images—e.g. mouth features to categorize *face expression*; left/right eye features for *face gender*; body parts for *pedestrian gender*; different features for *vehicle type*. Importantly, different participants often use different features to categorize the same object using the same labels—e.g. Figure 1C illustrates that participant 1 uses the windshield and a large portion of the front fender and bonnet to classify *vehicle type* as “sedan” or “SUV” whereas participant 10 uses the shape of the alloy wheel and the vehicle badge on the hood to produce the same category labels (Figure S1 for per participant results). This demonstrates the key, but often neglected, point that a similar stimulus-response relationship across participants (or participant and models) does not warrant internal processing of the same stimulus features[2].

Having identified these categorization features in each participant, we can now uniquely examine how their brain transforms identical high-dimensional stimulus images into participant-and-task-specific low-dimensional feature manifolds to enact task-behavior.

**Brain: Systems-level time-courses of task-dependent stimulus transformations**

To identify the stages that transform the high-dimensional input images into the low-dimensional feature manifolds underlying categorization behavior, we used a data-driven analysis. This analysis computed the representation of each varying scene pixel across trials (due to random Bubbles sampling) into the corresponding variations of MEG source amplitude responses post-stimulus–i.e. computing MI(pixel-visibility, $MEG_t$), for each pairing of 61 x 47 pixels and 5,107 cortical sources, across 271 time points. We segmented the results into brief consecutive periods covering key neural events involved with visual categorizations–i.e. C1,[52] occipital hemifield responses; P100, early attention/stimulus representation and ensuing N170,[53–57] faces/familiar object representations; N250[58] and P300,[45] attention/decision mechanisms.

Figure 2A summarizes the results, showing different dynamic transformations of the same stimulus images in each task (rows), where orange-to-yellow colors indicate number of participants whose sources represent this image pixel in each period; maximum = 9/10 participant, Maximum a Posteriori Probability (MAP) estimate of the population prevalence[48] of the effect of 9/10 participant replications = 0.9 (95% highest posterior density interval HPDI,[48] [0.61 - 0.99]), see *Methods, Analyses, Global representation of image pixels in brain networks*; *Methods, Bayesian population prevalence.*

Considering each pixel as a stimulus dimension, each task shows MEG sources transitioning from an initially high-dimensional stimulus representation of large parts of the scene (Stage 1, 50-120 ms, periods 1 to 4) to a more focused representation of only the task-relevant pixels–i.e. the lower-dimensional feature manifolds that develop between Stage 3, periods 7 to 9, 161-350 ms, compare with Figure S1. Stage 2 (periods 5 to 6) therefore marks the critical transition from higher-dimensional Stage 1 to task-relevant feature manifolds Stage 3.

To formalize these transitions, we grouped image pixels as either task-relevant or irrelevant based on participant behavior–cf. Figure 1C and Figure S1 and *Methods, Analyses, Feature mask and visibility*. In Figure 2B, the red curve shows across different periods the number of task-relevant pixels; the blue curve shows task-irrelevant ones. The cross-over of these curves between Stages 1-2 and Stage 3 identifies the transition from high-dimensional representations to task-specific feature manifolds.

**Brain: Systems-level localizations of task-dependent stimulus transformations**

To examine how localized MEG sources represent and transform the images depending on tasks, we compared how they represent an identical feature when it is task-relevant, or not– e.g. Participant 1's blue mouth in Figure 1C in *face expression* vs. in all other tasks. For each categorization feature (Figure S1), we therefore determined its per-trial visibility score *F*, by intersecting this feature's pixels with the pixels randomly sampled by Bubbles on each trial (cf. Figure 1, and *Methods, Analyses, Feature visibility*). For each participant, we then quantified the representation of *F* variations across trials[50,51] into corresponding of MEG variations of source amplitude responses–i.e. as MI(*F*; $MEG_t$ ),[50] FWER $p < 0.01$ over sources and time points, see *Methods, Feature representation on MEG sources.* Critically, the resulting MI curves (further developed in Figure 3) are **not** curves of brain activity. Rather, they are a proper measure of the representational strength of feature *F* into source activity ($MEG_t$).

First, Figure 2C shows summary results revealing where (which sources) and when (which Stage/time period) features are transformed when they are task-relevant and used for behavior vs. task-irrelevant. Each glass brain displays the number of participants (gray levels) whose MEG sources represent at least one such feature as task-relevant or irrelevant–i.e. maximum = 8/10 participant, MAP [95% HPDI] prevalence[48] = 0.80 [0.49 - 0.96]. For reference, we also show how these MEG sources represent category information for at least one task, computed for instance for *vehicle type* as MI(sedan vs. SUV stimulus; $MEG_t$), FWER $p < 0.05$ over sources and time points, plotted again as number of participants, maximum = 8/10 participant, MAP [95% HPDI] prevalence[48] = 0.80 [0.49 - 0.96].

Figure 2C reveals that occipital cortex sources represent both task-relevant and task-irrelevant features during Stage 1, accounting for its higher dimensionality. However, the task effect is already present with early representations distributing around the image locations of task-relevant features (though MI effect sizes are weaker for surrounding pixels, Figure 2A). Task-relevant features move along the ventral/dorsal pathways, but when irrelevant, they halt at the occipital-temporal junction. Occipital sources reduce most task-irrelevant features while ventral sources form a lower-dimensional feature manifold during Stage 2.[i] During Stage 3, the occipito-ventral/dorsal pathways keep transforming the low-dimensional feature manifolds into task-relevant features (Stage 3, period 9). Category information is increasingly represented from 161-280 ms (Stage 3, period 7 and 8), peaking at parietal-frontal juncture post ~291 ms (Stage 3, period 9).

**Brain: Systems-level expansion of task x feature transformations**

Figure 3 expands Figure 2C, displaying representational dynamics in a feature x task grid. Each panel shows the cross-participant average curves of significant feature representation–i.e. MI($F$; $MEG_t$)–every 1.67 ms, for each color-coded MEG source, progressing from cyan (occipital) to yellow (frontal).

[i] Figure S2 develops the analysis shown in Figure 2A to demonstrate that the dynamic transformations of broader image representations into lower-dimensional manifolds across the posterior-anterior axis of the ventral pathway coincide over Stage 2 with the reduction of task-irrelevant features in occipital lobe.

The diagonal of Figure 3A (box highlights) shows task-relevant features (*Fs*) transformations through Stage 1 to 3 (dashed lines), with representations progressing from occipital to higher-level regions (cyan-to-yellow occipital-to-frontal time-courses). Off-diagonal plots display short-lived representations of task-irrelevant features confined to occipital sources (cyan curves). Small glass brains above the representation curves localize the sources generating task-relevant vs. irrelevant feature representations–grey scale indicates number of participants with significant representation (MI) of the specified feature in each source, during Stage 1-2 (left) and Stage 3 (right). An exception is the central face's eyes, which remain represented in the two face tasks, consistent with previous studies.[56,59,60] This observation will be revisited in the Discussion.

In Figure 3A, comparison between features when they are task-relevant (diagonal) vs. irrelevant (each row, off-diagonal) is noteworthy, showing similar initial occipital (cyan) representations of the same feature. However, by Stage 2, these representations diverge, with the same feature reduced in occipital lobe vs. passed into ventral/dorsal pathways. Next, we develop the mechanisms behind this divergence at the level of individual occipital sources.

**Brain: Source-level representation of task-relevant vs. irrelevant *F***

First, we draw attention to the higher-dimensional feature representations on cyan occipital sources (Figure 3A) which align at Stage 1 with the peak cross-trial variance of their evoked MEG responses (Figure 3B). During 120-150 ms Stage 2, this variance drops (negative gradient in 3B), marking the time when occipital sources reduce task-irrelevant features while relevant features progress into ventral/dorsal pathways. Figure S3 further shows that occipital representations of a given feature peak slightly sooner when it is task-irrelevant than task-relevant (Wilcoxon rank sum test, $p < 0.001$, MI averaged at each time point across participants, tasks and sources). Now, we investigate how the variance of an occipital source marks the identical feature variations $F$ as task-relevant (passed for further processing) or task-irrelevant (reduced).

Figure 4A illustrates the variations $F$ of the vehicle feature visibility from participant 8 as disks with varying radii. Figure 4B shows the representation curve of $F$ on a cyan occipital source at Stage 1, when the feature is relevant (*vehicle type* task, Figure 4B, solid cyan MI representation curve for this source) vs. irrelevant (*all other tasks*, Figure 4B, dashed cyan

MI curve). Figure 4C shows corresponding ERF and variance underpinning these representations of *F*.

Figure 4D shows how the occipital source differently represent vehicle feature *F* based on its task-relevance. At Stage 1 (111 ms post-stimulus), identical variations of *F* (disk radii) exhibit an opposite representational relationship with MEG amplitude responses on the source. Critically, this depends on whether *F* is task-relevant (solid arrow in 4D), and subsequently passed into the ventral/dorsal pathways, or task-irrelevant (dashed arrow), and subsequently reduced in occipital cortex (i.e *vehicle type* vs. all other tasks). Figure 4E quantifies such opponent representations with information theoretic synergy(*F*; $MEG_t$; task-relevance vs. task-irrelevance) that Figure 4D illustrates at its 111 ms peak (indicated with opponent cyan arrows in Figure 4E cyan curve). Task-synergy quantifies how the same feature is differently represented on the same source[ii] depending on task. We consistently found such opponent sources (Figure 4G, opposite cyan arrows) in occipital cortex during Stage 1 (FWER $p < 0.01$ over MI-significant sources * 271 time points, see *Methods, Analyses, Opponent feature representations*).

In contrast, task-synergy can also indicate a feature that is unidirectionally represented either when it is task-relevant or irrelevant. Figure 4E displays the synergy curve of an example right Fusiform Gyrus (rFG) source, marked in purple, with a 135 ms peak (single arrow on the curve) during transition Stage 2. In Figure 4F, this rFG source unidirectionally represents the same vehicle feature *F*, but here only when it is task-relevant. Figure 4G extends this observation, illustrating across sources and time the count of participants who have at least one such exclusive task-relevant (or task-irrelevant) feature representation (FWER $p < 0.01$ over MI-significant sources * 271 time points, see *Methods, Analyses, Task-relevant feature representation*).

**Brain: Network interactions modulate early source representations by task**

Figure 4 shows that amplitude variations in occipital sources can represent the same feature differently by task relevance: either in opposite directions (opponent sources) or unidirectionally (unidirectional sources). And the direction of amplitude responses at Stage 1 could determine whether the feature will be reduced at Stage 2 or prominently represented for behavior in Stage 3. Here, we test whether network interactions between Pre-Frontal Cortex (PFC) regions and the occipital-ventral/dorsal pathways during Stages 1 and 2 top-

---

[ii] Although the sign of MEG responses is arbitrary[61], opponent signs reliably indicate the task-relevance vs. irrelevance of the same feature.

down modulate these early feature representations when they are task-relevant vs. task-irrelevant.

To investigate this, in each participant and task, we pinpointed two sources in the occipito-ventral/dorsal pathway during Stages 1 and 2: that with strongest opponent representation of a given feature *F* (the synergistic “opponent seed” shown in Figure 5, color-coded by participant) and that with strongest unidirectional representation of *F* (the synergistic “unidirectional seed” also shown in Figure 5). We then computed separately how opponent and unidirectional seeds interact with all PFC sources–i.e. as synergy(*F*; seed $\text{source}_t$; PFC $\text{source}_t$), separately for trials when *F* is task-relevant vs. *F* is task-irrelevant, FWER $p < 0.05$. Synergy emerges when two sources together predict more information about the feature than the sum of prediction by each source. Though PFC brain activity does not directly represent the feature, it does influence representation of the feature in the occipital-ventral/dorsal pathways. That is, when PFC activity changes, the relationship between feature visibility and activity in occipital-ventral/dorsal pathways changes. In this case, PFC sources and occipital-ventral/dorsal sources will generate synergy (the extra information about the feature that cannot be obtained from only occipital-ventral/dorsal sources without considering the PFC). Therefore, we need to explicitly consider PFC and occipital-ventral/dorsal activity together (synergistically) to understand the role of PFC on the occipital-ventral/dorsal representation of the feature.

This synergy analysis revealed the four spatio-temporal maps in the PFC shown in Figure 5–i.e. opponent and unidirectional seeds x task-relevant and irrelevant feature conditions. Each map indicates where, when and how strongly each pair of PFC and occipito-ventral/dorsal sources worked together as a network in representing feature *F*, separately for when *F* was task-relevant and irrelevant.

When *F* is task-relevant, Figure 5A shows that the orbitofrontal and ventromedial PFC (vmPFC) interact with both unidirectional and opponent seed sources during Stages 1 and 2 (96-150ms)–i.e. unidirectional seeds, maximum = 8/10 participant, MAP [95% HPDI] prevalence[48] = 0.80 [0.49 – 0.96]; opponent seeds, maximum = 7/10 participant, MAP [95% HPDI] prevalence[48] = 0.70 [0.38 – 0.90]). Critically, vmPFC interacts with occipital opponent sources primarily during Stage 2, when occipital cortex passes task-relevant features into the ventral pathway but reduces task-irrelevant features. This suggests that vmPFC is involved with maintaining representations of stimulus features when they are task-relevant across Stages 1 and 2, enabling their subsequent processing in the ventral pathway.

In contrast, when the same *F* is task-irrelevant, Figure 5B shows that the unidirectional occipital sources interact primarily with PFC orbitofrontal region, in Stage 1 (71-95 ms); maximum = 9/10 participant, MAP [95% HPDI] prevalence[48] = 0.90 [0.61 – 0.99]. There is no clear PFC network interaction pattern for opponent seeds. These network interactions suggest that the PFC plays a role in guiding early attention and feature reduction. The orbitofrontral region of PFC interacts with the occipital-ventral/dorsal pathway to represent task-irrelevant features, before these features are halted at the junction between the occipital and temporal regions, followed by their subsequent reduction during Stage 2.

In sum, our network analyses show that different regions of PFC get involved with the early occipito-ventral/dorsal representations of the same stimulus feature, depending on its task-relevance. Specifically, when a feature is task-relevant, orbitofrontal PFC and vmPFC modulate its unidirectional and opponent representations during Stages 1 and 2 (~96-150 ms), when the feature progresses from occipital into ventral/dorsal pathways for processing for behavior. In contrast, when the same physical feature is task-irrelevant, orbitofrontal PFC modulates its unidirectional occipital representation at Stage 1 (~71-120 ms), that occipital cortex then reduces from ~120 ms. These distinct network interactions therefore suggest that PFC regulates how occipito-ventral/dorsal pathways transform image representations into the lower-dimensional feature manifolds based on the task at hand.

## Discussion

At a systems-level, we aimed to provide a detailed descriptive model of where, when and how the brain networks of individual participants transform an identical set of high-dimensional input images into different low-dimensional manifolds of categorization features that support behavior in four different tasks—i.e. *face expression*, *face gender*, *pedestrian gender* and *vehicle type*. We revealed three stages that transform stimulus features into task-specific manifolds under pre-frontal cortex influence. Using precision neuroimaging and a dense-sampled design, we replicated these three stages in at least 8/10 participants, conferring high Bayesian population replication probability.[48]

Feature processing[62,63] is foundational to recognition,[64,65] working[66–68] and semantic memory,[69,70] extending to conscious perception.[71,72] and now crucial to understand interactive hierarchical models that disambiguate representations across cortical layers.[14,15,38,73] For instance, categorizing *vehicle type* should elicit predictions of the participant's vehicle features, whose top-down flow to occipital cortex should interact with bottom-up input. We showed that distinct tasks (e.g. *vehicle type* vs. *pedestrian gender*) elicited top-down PFC influences from Stage 1, suggesting that network mechanisms determine relevance (and progression) vs. irrelevance (and reduction) of the same physical feature from occipital cortex. Categorization models, including Deep Neural Networks (DNNs) should replicate these mechanisms, yielding similarly understandable feature manifolds in each task.[2] Otherwise, though DNNs might predict category membership as humans do, features and transformations might diverge.[74]

*The critical first 150 ms*

Stimulus transformations observed at Stages 1 and 2 largely align with early selection models of attention.[75,76] Task-relevant features are selected (filtered in) and transformed for behavior whereas task-irrelevant features are quickly reduced (filtered out). Synergistic interactions during Stage 1 (orbito-frontal and ventral-medial PFC) orchestrate feature processing based on task-relevance vs. irrelevance. The synergistic interactions mean that Stage 1 PFC activity modulates representation of features in occipito-ventral/dorsal pathways rather than directly responding to their visibility. This aligns with an early top-down modulation of PFC on the bottom-up feature representations leading to behavior,[17,77] underscoring the role of task constrains on feature filtering mechanisms in capacity-limited systems.[78] Further research delving into finer granularities of neural responses,[16,79–81] could inform these mechanisms.

Sustained feature manifold representations occur in each task between stimulus and behavior. How do gain functions and recurrent/interactive activations in the pathways' cortical layers uphold these task-relevant feature representations? Conversely, when these same features are task-irrelevant, they are only briefly represented in occipital cortex. Fusing individual MEG source amplitude data with 7TfMRI cortical layer bold responses could provide insights in how the occipito-ventral cortical layers[41,80,82] differently represent identical stimulus features based on task-relevance. This could elucidate how opponent representations of the same feature, result from variations in layered cortical activity.

*The 100-170 ms occipito-ventral/dorsal junction and subsequent visual categorizations*

Task-dependent reduction vs. passing of stimulus features happen at the occipito-ventral and dorsal junction, around the timing of occipito-ventral N/M170 ERP.[53,55,56,83–86] The N/M170 reflects a network communicating to right fusiform gyrus stimulus features contra-laterally represented in occipital cortices.[84] Our results suggest a N/M170 reinterpretation. We showed that brain signal variance over ~50 ms preceding the N/M170 peak (Figure 3B) reflects when brain networks transition from high-dimensional (Stage 1) to lower-dimensional task-relevant feature manifolds (Stage 2). This transition could explain why the N170 has been associated with multiple face, object and scene categorizations.[53,55,83,84] Stage 2 transition also coincides with task-relevant features represented in occipital cortex converge/buffer into rFG.[4,84] Stage 3 could integrate[4,46] these lateralized features into bi-lateral representations (as suggested[42]). Here again, fusion of MEG and 7T fMRI,[16,80,82,87] could reveal how cortical layers integrate lateralized features into bilateral "stitched up" representations, pre- and post-170 ms, as we showed with simpler stimuli and tasks.[46,88]

When brain networks effectively categorize the stimulus relates to the feature-contents that are consciously accessed. Prevailing models[71,72] suggest features are "dispatched" to working memory for conscious access. Our data suggest feature manifolds are maintained from occipital to higher regions[84,89] jointly acting as functional memory.[72] Conscious access[90,91] could align with the manifolds at Stage 3, contrasting with features reduced in occipital cortex at Stage 2. This presents a tangible methodology to explore the complex landscape of conscious perception, including the influence of memory and prediction, as the feature manifolds likely represent predicted contents.[17,92]

Remember we flagged that the eyes were processed in both face tasks, even when irrelevant in *face expression*. We documented a similar result over the N170 ERP time-course[56,59] where the systematically represented eyes were not always necessary categorize facial expression. Others suggested the eyes are the first contact with a face.[93] We show that the brain represent the eyes and other face features in different tasks, which could explain why a more negative voltage N170 ERP is often reported for faces.[53,83,85] Such systematic rFG representations of spatially distributed features across the face could also explain its apparent "holistic representation."[94]

The image is more broadly represented in the face tasks than in the other two tasks. This likely results from a combination of inter-subject variability, cortical magnification, and attention. Specifically, the face is spatially broader in our stimuli than the pedestrian and vehicle. In the face tasks, participants likely attended to a larger region than in the pedestrian and vehicle tasks, which could also increase inter-subject variability. Together, these factors could have contributed to the broader image representation in the face tasks.

We studied pervasive mechanisms that dynamically transform the same complex, high-dimensional input images for multiple visual categorizations. Within 150ms post-stimulus, the occipital cortex, under frontal guidance, either passes or reduces a feature based on its relevance in a categorization task, revealing opponent representational signatures at the MEG source-level. Following this, occipito-ventral and dorsal networks focus on the feature manifolds relevant to each categorization task. These feature transformations offer mechanistic insights into attention theories, face and object categorizations, and our understanding of conscious perception.

## Acknowledgments

PGS was supported by the EPSRC [MURI 1720461] and the Wellcome Trust [107802]. PGS is a Royal Society Wolfson Fellow [RSWF\R3\183002]. RAAI was supported by the Wellcome Trust [214120/Z/18/Z].

## Author contributions

Conceptualization, Y.D., J.Z., J.G., R.A.A.I. and P.G.S.; data curation, J.G.; formal analysis, Y.D. and R.A.A.I.; investigation, Y.D., J.Z., R.A.A.I and P.G.S.; methodology, Y.D., R.A.A.I. and P.G.S.; software, R.A.A.I.; writing – original draft, Y.D. and P.G.S.; writing – review & editing, Y.D., J.Z., J.G., R.A.A.I. and P.G.S.; supervision, R.A.A.I. and P.G.S.; funding acquisition, P.G.S.

## Declaration of interests

The authors declare no competing interests.

## Supplemental information

Document S1. Figures S1-S6 and Tables S1-S2.

## Figures

**Figure 1. Categorization design and task-relevant features**
(A) Scene Images. 64 base images of a street scene comprise a central face flanked by a pedestrian and a parked vehicle.
(B) Randomly Sampled Features. On each trial, Bubbles randomly sampled pixels from one base image to synthesize a sampled stimulus. We used the same sampled stimuli[27] (presented in a random order) in each categorization task, so that each participant (N = 10) saw each sampled image 8 times (twice per task).
(C) Categorization Behavior; Task-Relevant Features. The stimuli afforded four different two-alternative forced choice categorizations: *face expression*, “happy vs. neutral” responses; *face gender*, “male vs. female”; *pedestrian gender*, “male vs. female”; *vehicle type*, “sedan vs. SUV.” *Task-relevant features* For each participant and image pixel, we computed MI(pixel-visibility; correct vs. incorrect categorization)[50] to reveal pixels that significantly (FWER $p < 0.05$) modulate categorization accuracy–color-coded for participants 1 and 10 to illustrate that participants often use different features to produce the same categorization responses (Figure S1 for all participants).

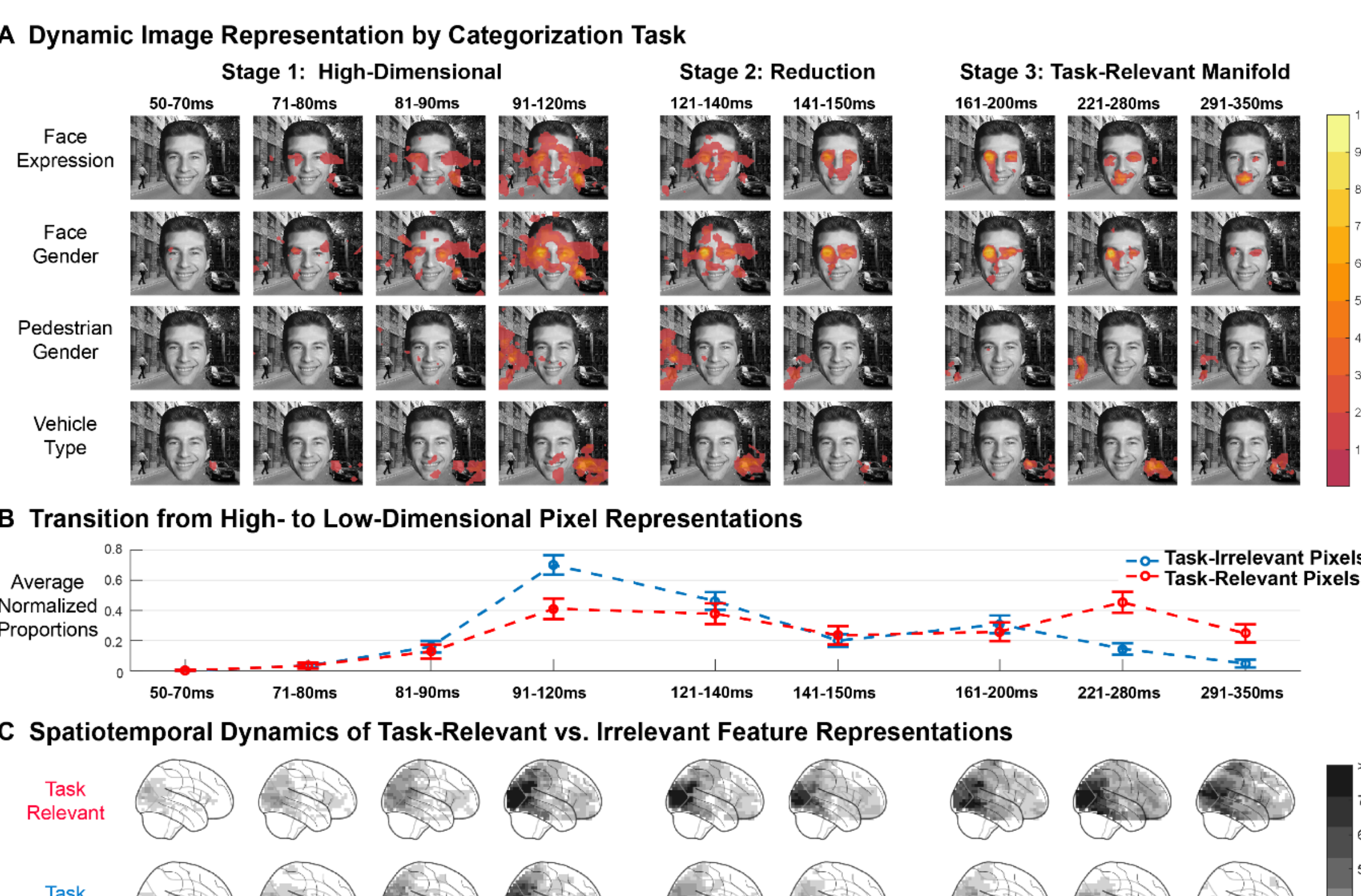

**Figure 2. Systems-level transformations of input images into categorization feature manifolds**
(A) Dynamic image representation by categorization task. In each participant and MEG source, we computed the cross-trial relationship between each pixel's visibility and the source amplitudes at time *t* post-stimulus–i.e. MI(pixel-visibility, $MEG_t$). To visualize how representations transform, we segmented the post-stimulus time course into 9 consecutive periods. In each, and for each categorization task, we pooled pixels significantly represented on at least one MEG source-False Discovery Rate (FDR) test with $q = 0.001$. Across participants we summarize results per period, revealing image pixels that participants' MEG sources represent in each task–orange-to-yellow colors indicate participant numbers whose MEG sources represent this pixel, maximum = 9/10 participant, MAP [95% HPDI] prevalence[48] = 0.90 [0.61 - 0.99]). Figure S2 develops the dynamic transformations across the posterior-anterior axis of the ventral pathway.
(B) Transition from high- to low-dimensional pixel representations. In each period, we computed across participants and tasks the average number of task-relevant (vs. task-irrelevant) pixels–i.e. normalized per participant and task to the maximum task-relevant (vs. irrelevant) pixel numbers, standard error bars provided. The resulting curves cross over between Stages 1-2 and Stage 3, showing transition from higher-dimension (comprising task-relevant and irrelevant pixels) to lower-dimensional feature manifolds (comprising task-relevant pixels).
(C) Spatio-temporal dynamics of task-relevant vs. irrelevant feature representations. For each participant, task and categorization feature (Figures 1C S1), we quantified how each source represents this feature (*F*) in its amplitude at each time point *t*–i.e. MI(*F*; $MEG_t$ ),[50] FWER $p < 0.01$. Greyscale sources show participant count that represented at least one feature in each period, when task-relevant vs. irrelevant, maximum = 8/10 participant, MAP [95% HPDI] prevalence[48] = 0.80 [0.49 - 0.96]. Figure S5 provides the glassbrain visualization with 4 projections. Category information provides a ground-truth reference of the MEG source representation of category information across participants—computed e.g. in *vehicle type* as MI(sedan vs. SUV stimulus; $MEG_t$), FWER $p < 0.05$, maximum = 8/10 participant, MAP [95% HPDI] prevalence[48] = 0.80 [0.49 - 0.96].

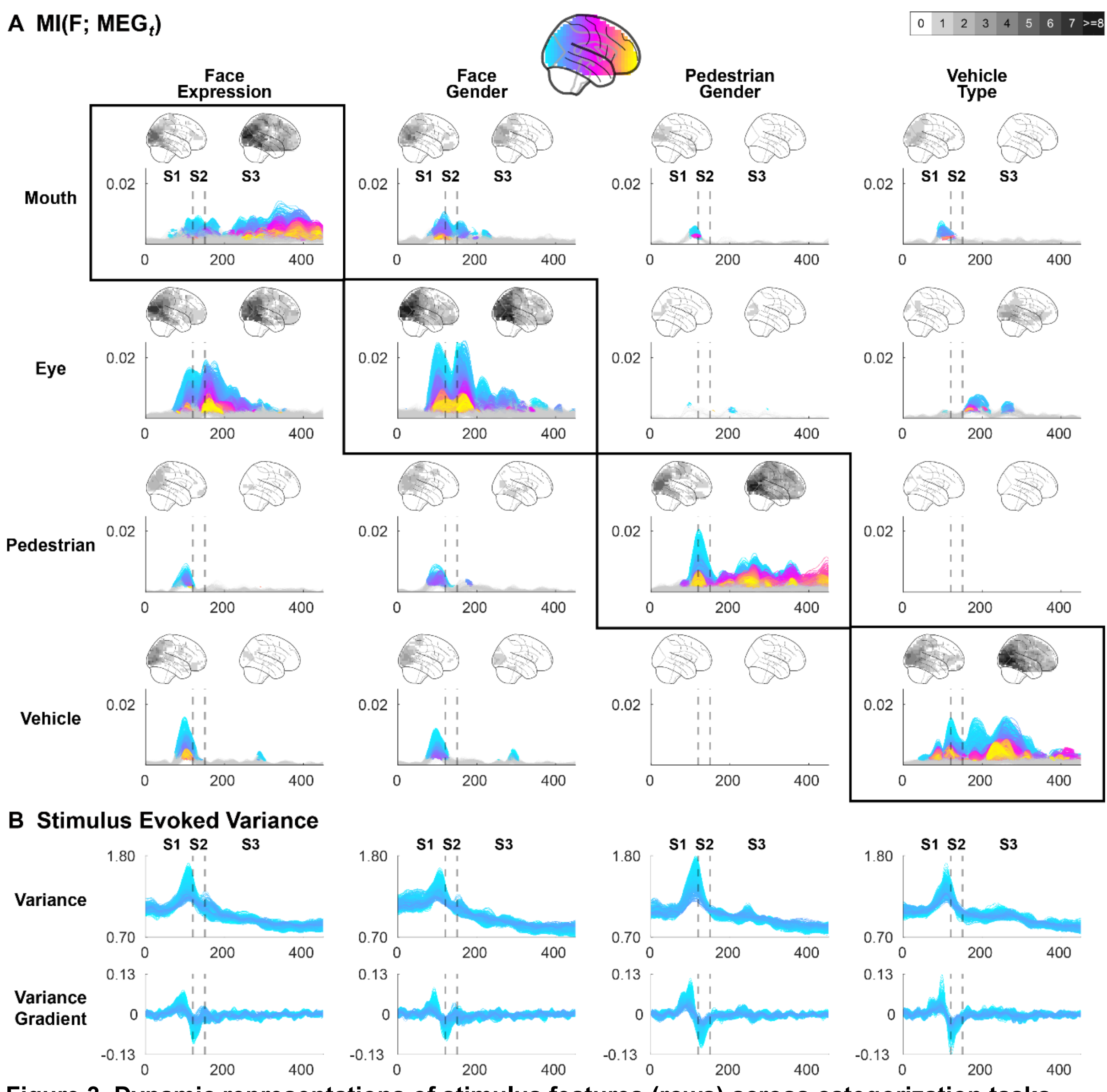

**Figure 3. Dynamic representations of stimulus features (rows) across categorization tasks (columns)**

(A) Curves in each cell show the participant average (n = 10) time-course of significant representation of each participant's feature (Figures 1C and S1) on MEG sources, computed as MI($F$; $MEG_t$),[50] FWER $p < 0.01$, each color-coded by its location on a posterior-to-anterior axis (cyan-yellow). Dashed lines (120 and 150 ms) delineate stages S1 to S3 reported from Figure 2. Small brains flanking the dashed lines[53,55,83,84] show the participant count whose MEG sources represent this feature 50-150 ms (left brain) and 150-450 ms (right brain) post-stimulus. Each row reveals qualitatively different representation dynamics of the same stimulus feature when task-relevant (matrix diagonal, box highlight) vs. task-irrelevant (off diagonal). Figure S3 further shows that occipital representations of a given feature peak slightly sooner when it is task-irrelevant than task-relevant (Wilcoxon rank sum test, $p < 0.001$).

(B) Stimulus evoked variance and gradient of MEG occipital source signal (cyan colored) in stages S1 to S3, averaged per source across participants.

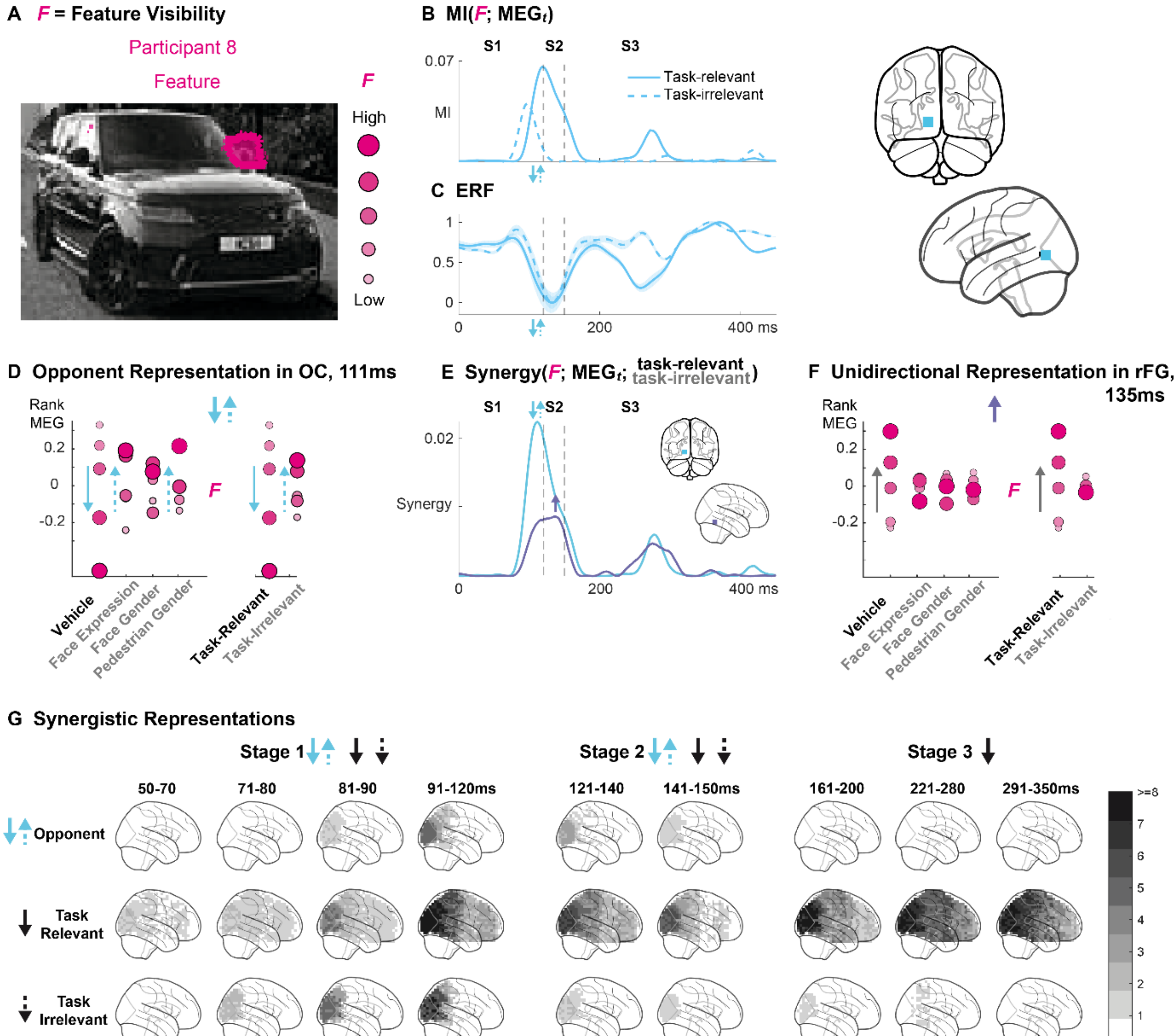

**Figure 4. Task-modulations of feature representations**

(A) Feature visibility, F. Random pixel sampling across trials varies visibility of vehicle feature *F*, represented as 5 varying radii for reference in panels D and F (participant 8, Figure S1).

(B) MI(F; MEG) quantifies dynamic representation of *F* in the occipital source located in small brains when *F* is task-relevant (plain cyan curve) vs. task-irrelevant (dashed cyan curve).

(C) Normalized Event-Related Field (ERF) underlying the MI curves (panel B) for this source whose MEG amplitudes variations (shaded area, variance) represent *F*.

(D) Opponent occipital source representations of F. At 111 ms, MEG amplitude variations of this same source (y axis) differently represent identical of feature *F* variations (circle radii, panel A) when task-relevant vs. irrelevant. Cyan arrows indicate these opposite directions when *F* is task-relevant (plain arrow, in *vehicle type*) and passed later into rFG vs. irrelevant (dashed arrow, other tasks) and reduced in occipital cortex. Figure S4 further illustrates the opponent representations.

(E) The cyan synergy curve quantifies the time course of these opponent representational interactions[50,51] (that panel D illustrates at peak 111 ms, indicated with opponent arrows). The dark blue synergy curve illustrates another representational interaction in the rFG source shown in panel F (located in adjacent small brain).

(F) Unidirectional Representations. Dark blue rFG source represents *F* at 135 ms peak synergy, but here only when the feature is relevant in *vehicle type*.

(G) Synergistic Representations. Synergy(*F*; $MEG_t$; Task-relevance) quantifies how brain sources differently represents identical *F* over Stages 1 to 3, covering three types of source-level representations. *Opponent synergy* indicates number of participants whose sources show significant opponent representations of the same *F* when task-relevant vs. irrelevant (cf. panel D); *Task-relevant (or irrelevant) synergy* indicates unidirectional representations of *F* when either task-relevant (cf. panel F) or irrelevant.

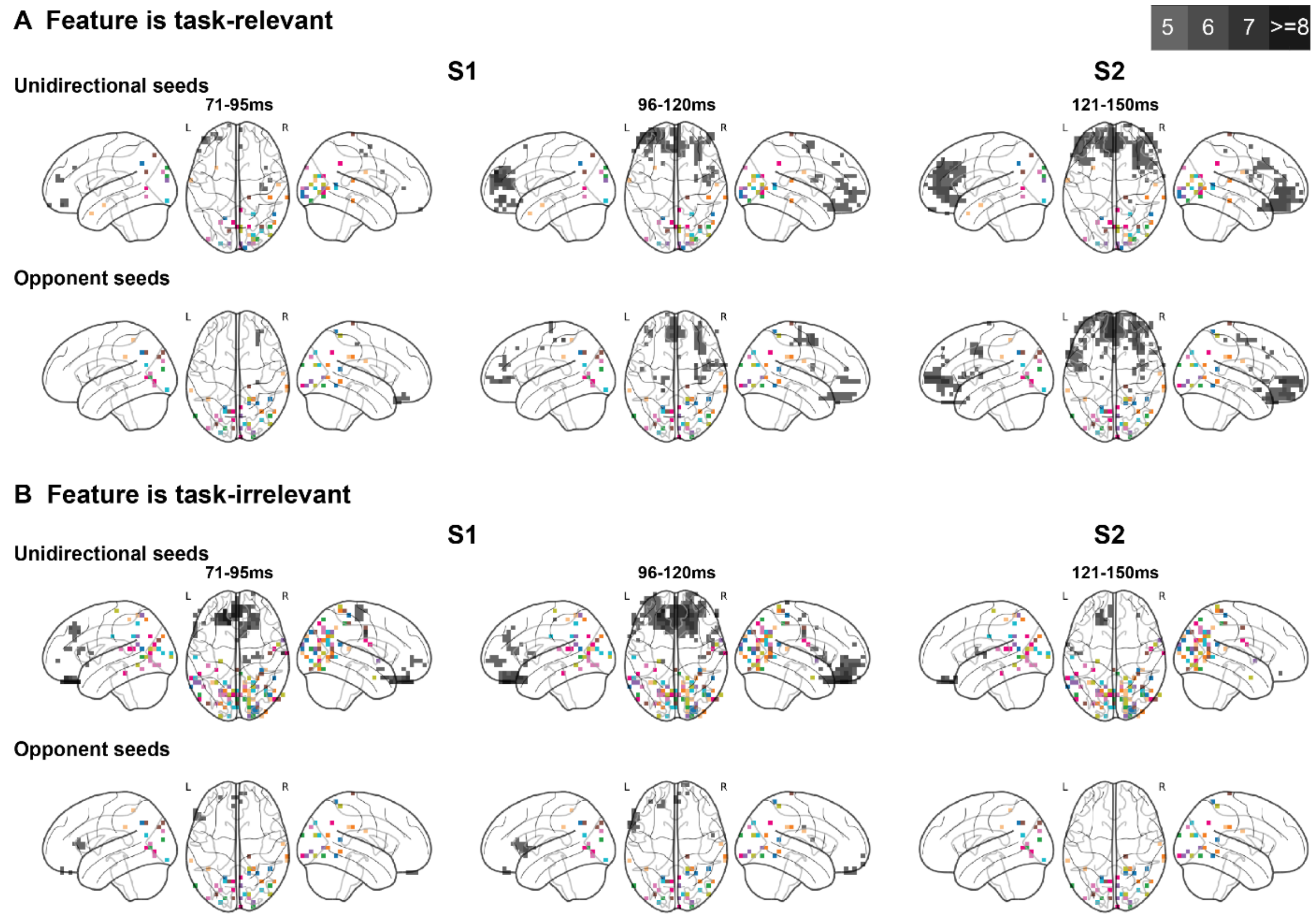

**Figure 5. Early network interactions between PFC sources and occipito-ventral/dorsal sources**
(A) Synergistic interactions when feature is task-relevant. Unidirectional and opponent occipito-ventral/dorsal seed sources are color-coded by participant. Grey-levels indicate participant prevalence (>= 5) of synergistic interactions, computed as synergy($F$; seed source$_t$; PFC source$_t$), revealing involvement of orbitofrontal and ventromedial PFC regions, from 96-120ms, FWER $p < 0.05$. Unidirectional seeds, maximum = 8/10 participant, MAP [95% HPDI] prevalence[48] = 0.80 [0.49 – 0.96]; Opponent seeds, maximum = 7/10 participant, MAP [95% HPDI] prevalence[48] = 0.70 [0.38 – 0.90].
(B) Synergistic interactions when feature is task-irrelevant. Unidirectional and opponent seeds synergistically interact mainly with orbitofrontal regions of PFC from 71-95ms, ending before the beginning of Stage 2 (121 ms). Unidirectional seeds, maximum = 9/10 participant, MAP [95% HPDI] prevalence[48] = 0.90 [0.61 – 0.99]; Opponent seeds, maximum = 6/10 participant, MAP [95% HPDI] prevalence[48] = 0.60 [0.28 – 0.83].

## Tables

**Table 1.** Cortical sources categorized into four regions of the Talairach-Daemon atlas.[95]

| Occipital region | Lingual gyrus (LG)<br>Cuneus (CUN)<br>Inferior Occipital Gyrus (IOG)<br>Middle Occipital Gyrus (MOG)<br>Superior Occipital Gyrus (SOG) |
|---|---|
| Temporal region | Fusiform Gyrus (FG)<br>Inferior Temporal Gyrus (ITG) |

| | Middle Temporal Gyrus (MTG)<br>Superior Temporal Gyrus (STG) |
|---|---|
| Parietal region | Superior Parietal Lobule (SPL)<br>Inferior Parietal Lobule (IPL)<br>Angular Gyrus (ANG)<br>Supramarginal Gyrus (SMRG)<br>Precuneus (PRECUN)<br>Postcentral Gyrus (POSTCEN) |
| Frontal region | Anterior Cingulate (AC)<br>Inferior Frontal Gyrus (IFG)<br>Medial Frontal Gyrus (MeFG)<br>Middle Frontal Gyrus (MiFG)<br>Orbital Gyrus (OG)<br>Paracentral Lobule (PL)<br>Precentral Gyrus (PRECEN)<br>Superior Frontal Gyrus (SFG) |

**Table 2.** Bayesian population prevalence: Maximum A Posteriori (MAP) [95% Highest Posterior Density Interval (HPDI)] for k significant participants out of 10.

| | Within participant α = 0.05 | Within participant α = 0.01 |
|---|---|---|
| K=10 | 1 [0.75 - 1] | 1 [0.75 – 1] |
| K=9 | 0.89 [0.61 – 0.99] | 0.90 [0.61 – 0.99] |
| K=8 | 0.79 [0.49 – 0.96] | 0.80 [0.49 – 0.96] |
| K=7 | 0.68 [0.38 – 0.90] | 0.70 [0.38 – 0.90] |
| K=6 | 0.58 [0.28 – 0.83] | 0.60 [0.28 – 0.83] |
| K=5 | 0.47 [0.19 – 0.75] | 0.49 [0.19 – 0.75] |
| K=4 | 0.37 [0.11 – 0.66] | 0.39 [0.11 – 0.66] |
| K=3 | 0.26 [0.05 – 0.56] | 0.29 [0.05 – 0.56] |
| K=2 | 0.16 [0 – 0.44] | 0.19 [0 – 0.44] |
| K=1 | 0.05 [0 – 0.34] | 0.09 [0 – 0.34] |

## STAR★Methods

## RESOURCE AVAILABILTY

### Lead contact

Further information and requests for resources should be directed to and will be fulfilled by the Lead Contact, Philippe G. Schyns (philippe.schyns@glasgow.ac.uk).

### Materials availability

This study did not generate new unique reagents.

### Data and code availability

Data reported in this study and the custom code for analyses are deposited at https://doi.org/10.17632/fd2zjrfgbc.2. Custom code for experiment and visualization are available by request to the Lead Contact.

## EXPERIMENTAL MODEL AND STUDY PARTICIPANT DETAILS

### Participants

Ten participants (3 males and 7 females, age: M = 25.3, SD = 1.64, range = 23-28 years old) with normal or corrected to normal vision participated in all four tasks and gave informed content. All participants were right-handed. Participant gender was not considered in the study design. The experiment was conducted in University of Münster, Germany. The study was approved by the ethics committee of the University of Münster (2019-198-f-S) and conducted in accordance with the Declaration of Helsinki.

## METHOD DETAILS

### Stimuli

We used 64 base greyscale images (8 face identities with 4 male and 4 female × 2 expressions × 2 pedestrians × 2 vehicles) of a realistic city street scene comprising the combinations of varying embedded targets: a central face (which was male vs. female and happy vs. neutral), left flanked by a pedestrian (male vs. female), right flanked by a parked vehicle (sedan vs. SUV). The images were presented at 5.72° × 4.4° of visual angle, with 364 × 280 pixel size. We sampled information from each image, using the Bubbles procedure. Specifically, we multiplied the image with randomly positioned Gaussian apertures (sigma = 15 pixels) to vary the visibility of image features on each trial. We used 35 Gaussian apertures in all tasks, which was determined by a behavioral experiment pilot with 4 participants to achieve strong categorization performance across the four tasks. Using 0.2 as the threshold for visible pixels, there were ~33% pixels preserved on average. However, the pixel visibility is a continuous scale instead of a simple binary state. Figure 1B provides two intuitive examples of the images showing how much of each image is preserved in each trial.

We pre-generated 768 random bubble masks which were the same in all categorization tasks. On each session of trials, we applied the 768 masks to 12 repetitions of the original 64 images, for a total of 768 trials presented in a random order.

**Task procedure**

Each trial began with a fixation cross presented for a random time interval 500-1000 ms, followed by one of the original stimuli for 150 ms, whose features were randomly sampled with the Bubbles procedure. Participants were instructed to maintain fixation on each trial and respond as quickly and accurately as possible, by pressing one of two keys ascribed to each response choice—i.e. "happy" vs. "neutral" in *face expression*; "male" vs. "female" in *face gender* task; "male" vs. "female" in *pedestrian gender*; "sedan" vs. "SUV "in *vehicle type*. Each task comprised two sessions of trials, each comprising 768 trials (of 6 runs followed by a short break, each run comprising 128 trials = 8 identities × 2 expressions × 2 pedestrians × 2 vehicles × 2 repetitions). Tables S1 and S2 report the categorization accuracy and reaction time.

**MEG data acquisition**

Participants were seated upright in a magnetically shielded room while we simultaneously recorded MEG and behavior data. Brain activity was recorded using a 275 channel whole-head MEG system (OMEGA 275, VSM Medtech Ltd., Vancouver, Canada) at a sampling rate of 600 Hz. During MEG recordings, head position was continuously tracked online by the CTF acquisition system. For MEG source localization, we obtained high-resolution structural magnetic resonance imaging (MRI) scans in a 3T Magnetom Prisma scanner (Siemens, Erlangen, Germany).

**Pre-processing**

We performed analyses with Fieldtrip[96] and in-house MATLAB code, following recommended guidelines.[61] We first visually identified noisy channels and trials with epoched data (-400 to 1500 ms around stimulus onset on each trial) high-pass filtered at 1 Hz (4th order two-pass Butterworth IIR filter). Next, we epoched the raw data into trial windows (-400 to 1500 ms around stimulus onset, 1-25 Hz band-pass, 4th order two-pass Butterworth IIR filter), filtered for line noise (notch filter in frequency space), applied fieldtrip built-in denoise function specific to the MEG system, and rejected noisy channels and trials identified in the first step. We then decomposed the data with ICA, and visually identified and removed the independent component corresponding to artifacts (eye blinks or movements, heartbeat).

**Source reconstruction**

We applied a Linearly Constrained Minimum Variance (LCMV) beamformer[97] to reconstruct the time series of 12,773 sources on a 6mm uniform grid warped to standardized MNI coordinate space. Using a Talairach-Daemon atlas,[95] we excluded all cerebellar and non-cortical sources, and performed statistical analyses on the remaining 5,107 cortical grid sources. We categorized cortical sources into four regions based on ROIs defined in the Talaraich-Daemon atlas.[95] Figure S6 visualizes the localization of these regions.

## QUANTIFICATION AND STATISTICAL ANALYSIS

**Feature representation**

*What is it?*

Feature representation refers to a systematic relationship between a feature of the external world and neural activity.[98] Our methodology quantifies the representation of a visual feature so that we can trace where, when and how the brain processes it.

*How is a feature representation quantified?*

In our data, the visibility of a feature in a stimulus varies in a continuous manner across trials–i.e. it is not a binary feature present vs. absent. To measure the representation of the feature into MEG activity, we use Mutual Information (specifically, the Gaussian Copula MI, GCMI).[50] GCMI quantifies across trials how strongly the variations of MEG amplitude represent the variations of feature visibility in the stimuli–i.e. as the information that MEG amplitude variations and feature visibility variations share, measured on the scale of bits.

For example, Figure S4A now plots the mean MEG amplitude response curves, where all trials are split into 5 equally occupied feature visibility bins–quintiles of the empirical CDF of feature visibility. Statistical difference between these mean curves is considered to reflect important processing differences across feature visibility conditions. Figure S4 clarifies that the highest MI measure of feature representation corresponds to largest differences amongst mean MEG responses to the different bins of feature visibility.

The MEG amplitude curves evolve with peaks and troughs. These peaks and troughs can reflect representations of other features and/or cognitive variables. However, the feature representation curve underneath in Figure S4B does not mirror the MEG peaks because our information theoretic analysis specifically isolates, from raw MEG amplitude variations, the information that only pertains to the tested stimulus feature.

**Feature manifolds**

We used 'manifold' in its mathematical understanding, as a topological space that locally resembles Euclidean space. In neuroscience, 'neural manifold' is often used to refer to geometric structures in neural population activity–i.e. a subspace of neural state space.[99,100]

We deliberately used 'feature manifold' to refer to the geometric structure of visual inputs (e.g. images) that are represented in neural activity. Object categorization relies on diagnostic features, which we show underline categorization behavior. However, a given object can be categorized in multiple different ways, each relying on distinct sets of diagnostic features. This implies that the brain must represent different stimulus feature manifolds for this object. This is often neglected in neuroimaging studies of visual categorization. We show that only a subspace of the 2D projection of the real-world (i.e. the image) is selected for categorization, in a task and participant-specific way.

**Participant features (supports Figures 1 and 2 and S1)**

To reveal what image features each participant used to in each categorization task (i.e. the task-relevant features), we quantified the cross-trial statistical dependence between the visibility of each pixel (due to bubbles sampling[27] on each trial) and the corresponding correct vs. incorrect response of the participant in this task, computed as Mutual Information,[50,51] MI(pixel visibility; correct vs. incorrect categorization). We represented pixel visibility on each trial as a real number from 0 to 1 (low to high visibility), which we then binarized using a 0.2 threshold into 2 categories: 0 for low visibility and 1 for high visibility. To establish statistical significance, we ran a non-parametric permutation test with 1,000 shuffled repetitions, corrected over 364 x 280 pixels using maximum statistics (FWER $p < 0.05$). Significant pixels represent the participant's task-relevant features whose visibility influences their categorization behavior in each task (see Figure 1, Figure 2A and Figure S1).

Generally, participants used similar strategies in two sessions, though some used different strategies. Because features obtained from reverse correlation cover different aspects of the image they are constrained to the same psychophysical scale between tasks. We highlight that this is not problematic for our rank-based information theoretic analysis.

**Global representation of image pixels in brain networks (supports Figure 2)**

To visualize the global representational dynamics of the visual stimuli in each categorization task, we computed MI(pixel-visibility, $MEG_t$) for each one of the 364 x 280 stimulus pixels (downsampled to 61 x 47 for computational efficiency), 5,107 cortical MEG sources and 271 time points, producing a 3D matrix of MI values of dimensions 2867 (61 x 47) pixels x 271 time points x 5,107 sources. MI quantifies the statistical dependence between two variables. We calculate MI between the continuous valued pixel visibility (bubble mask value) and the continuous valued MEG amplitude at a given source and timepoint with Gaussian Copula Mutual Information (GCMI).[50] The empirical Cumulative Distribution Function (CDF) of the marginal distribution of each variable (pixel visibility and MEG) is estimated, and the data are transformed via the inverse CDF of a standard normal distribution. This results in a data set with perfect standard normal marginal distributions, but the same empirical copula as the original data. Then standard analytic expressions for bias-corrected Gaussian MI are used. As MI is invariant to marginal distributions, and Gaussian distribution has maximum entropy given constrained second moments, this GCMI procedure provides a lower bound estimate of the true MI.[50]

We then segmented the time dimension into nine periods ([50-70], [71-80], [81-90], [91-120], [121-140], [141-150], [161-200], [221-280], [291-350] ms). To visualize the pixels that the MEG sources of each participant represent, we pooled all the pixels with statistically significant MI on at least one source on the considered period. To compute this statistical significance in each participant, for each pixel, we took the maximum MI(pixel visibility, $MEG_t$) at each time point, resulting in a pixels x time matrix. We performed a FDR test on this matrix with a false discovery rate set at $q = 0.001$. For each image pixel, we color-coded in Figure 2A the number of participants with such significant MI (maximum number = 9, Maximum A Posteriori (MAP) [95% Highest Posterior Density Interval (HPDI) prevalence[48] = 0.90 [0.61 – 0.99]). Figures deposited at https://doi.org/10.17632/fd2zjrfgbc.2 show image representational transformations over the entire time course.

**Feature mask and visibility (supports Figure 3 and 4)**

As different participants can use different features in each task, to generalize analyses across participants, we transformed the data from levels of pixel visibility into levels of feature visibility $F$ (i.e. comprising the pixels making up the features of each participant). To this end, for each feature we selected the top 5% pixels with highest MI(pixel visibility; correct vs. incorrect categorization) to form feature masks. On each trial, we computed feature visibility $F$ as the feature mask pixels shown by the bubbles sampling, weighted by the MI values of each pixel of the feature mask. To better trace the early contralateral projection in visual cortex, we divided mouth (for *face expression*) and eyes (for *face gender*) features into their left and right components and considered them as a 2-dimensional feature variable in our analyses. Figure S1 shows the feature masks of each participant and task.

$$F = \sum_i MI(Pixel_i\ visibility; Behavior) \cdot Pixel_i\ visibility \quad (1)$$

It is worth noting that the maximum statistic that we use is stringent. A participant without any significant pixel in the behavioral analysis does not necessarily imply that these pixels have no influence on their behavior or are not represented in their brain activity. Therefore, when calculating feature representations in the brain in following analysis, we used as the task-relevant feature the top 5% of pixels with highest MI. To eliminate the effects of noise pixels included in the top 5% pixels, we used MI(pixel visibility; MEG amplitude) as the weight for each pixel when calculating the feature visibility. Thus, the contribution of those pixels with low MI will be small.

**Feature representation on MEG sources (supports Figures 2, 3 and 4)**

To reconstruct where, when and how MEG sources represent each participant's features, we computed MI between *F* of each feature and 5107 MEG source signals over 0 to 450 ms, in each task—i.e. when the feature is task-relevant, and also in the three other tasks when it is task-irrelevant, computed as MI($F$; $MEG_t$) with GCMI as described above.[50] To establish statistical significance, we ran a non-parametric permutation test with 1,000 shuffled repetitions, corrected (FWER $p < 0.01$) over 5107 sources x 271 timepoints with maximum statistics. This computation produces a 4 (tasks) x 4 (features) x 5,107 sources x 271 time points feature representation matrix for each participant. See Figures deposited at https://doi.org/10.17632/fd2zjrfgbc.2 for the spatiotemporal dynamics of feature representations in individual participants.

**Variance of MEG amplitude**

We measured the cross-trial variability of the MEG amplitude at each time point. It was calculated by var(MEG amplitude) ./ mean(var(MEG amplitude)), where mean(var(MEG amplitude)) is the average variance of MEG amplitude over time for normalization. So, Figure 3B shows how the cross-trial variability of MEG amplitude changes over time. Gradient is rate of change or slope.

**Task modulation of feature representation on MEG sources (supports Figures 3 and 4)**

When considering the influence of the task factor on the representation of *F*, synergy comes into play. We use information theoretic synergy to quantify how the variables MEG response and task together provide more information than the sum of the information provided by each variable individually:

$$\text{Synergy}(F;\ \text{MEG}_t;\ \text{Task}) = \text{MI}(F;\ \text{MEG}_t \mid \text{Task}) - \text{MI}(F;\ \text{MEG}_t) \quad (2)$$

If the task doesn't influence the representation of *F* in MEG then the two mutual information (MI) terms in (2) won't show any difference, leading to a synergy value of zero. That is, the difference in the representation strength of *F* in MEG amplitude when we condition out the effect of task (first MI term) vs. when the effect of task is present (second MI term) will be 0. If the task factor does not influence *F* representation, then the MI quantities in (2) will not differ resulting in zero synergy. Thus, significant synergy occurs when the average representational strength of *F* is higher when the task is controlled, implying that we would better predict the *F* from brain activity if we also knew what task was performed. Synergy quantifies how much the different tasks modulate the feature representation in MEG.

To quantify the modulation effect of the four categorization tasks on the representation of the participant's features into MEG source activity, we computed information theoretic synergy, as just defined, between 0 and 450 ms post-stimulus, for each participant feature, where the categorization tasks variable has values of 1 to 4 to represent each task. To establish statistical significance, we use a nonparametric permutation test, with 1,000 repetitions, shuffling the task label of each trial, corrected over 5107 sources * 271 timepoints (FWER $p < 0.01$). This provides permutation samples from the null distribution where task does not affect feature representation.

**Task-relevant vs. task-irrelevant**

To quantify the specific modulation of task-relevance vs. irrelevance on the MEG source representation of each participant feature, we computed again synergy, this time as synergy($F$; $\text{MEG}_t$; task-relevance), where task-relevance could be 1 (for task-relevant) or 0 (for task-irrelevant). We observed synergy arising from two different representational mechanisms: Opponent feature representation and unidirectional feature representation. We define each below.

*Opponent feature representations*

Opponent feature representation on a given source means this: **the same physical variations of feature visibility incur MEG amplitudes in opposite directions depending on whether this feature is task-relevant vs. task-irrelevant.** Figure S4 illustrates this opposition in the shaded time window. We can see that the same changes in feature visibility give rise to MEG amplitude changes in opposite directions in the task-relevant and task-irrelevant binned MEG amplitude curves. Specifically, when the feature is task-relevant, the MEG amplitude response is more negative to higher feature visibility; in contrast, when the feature is task-irrelevant, the MEG amplitude response is more negative to lower feature visibility. It is important to note that this reversal refers to a difference in the sign of the correlation between feature and MEG–although we use MI, which is an unsigned measure. This reversal is **not** a statement about the sign of the evoked magnetic field. As shown in the example, there is a change in the sign of the correlation relationship, but the evoked MEG signal has negative sign in both cases. This implies that significant MI for *F* in multiple tasks, but their synergy reveals that this representational relationship depends on task.

We formalize this effect as the following logical conjunction (see Figure 4C).:

Opponent feature representation:-
<significant task-relevant MI>
& <task-irrelevant MI>
& <significant synergy>
& <opponent signs for relevant vs irrelevant>

*Unidirectional task-relevant feature representation*
Occurs when a given source represents a participant feature only when it is task-relevant. We define this synergy as (see Figure 4D):

Unidirectional task-relevant feature representation:-
<significant task-relevant MI>
& <no significant task-irrelevant MI>
& <significant synergy>

Given that eyes are represented in brain activity in the face expression task when they are task-irrelevant, we excluded the eyes in face expression task from the task-irrelevant condition.

**Bayesian population prevalence**

Table 2 below provides a reference to transform the proportion of participants from the sample who have a significant effect into the Bayesian population prevalence[48]. Population prevalence is a Bayesian estimate of the within-participant replication probability. Replicating a result in multiple participants offers a much higher standard of evidence than to declare statistical significance of a population mean effect. For example, $p = 0.05$ typically defines population mean statistical significance; $p < 0.001$ would be considered stronger evidence. In Figure 3A (diagonal of the matrix), we show 8/10 participants have significant MI task-relevant feature representations in occipital and ventral cortex (FWER $p < 0.01$). The p-value corresponding to this result (i.e. for the observation 8/10 significant at p=0.05) under the global null that no one in the population shows this effect is $1.6 \times 10^{-9}$. Under the global null our results are therefore 7 orders of magnitude more surprising than a typical mean demonstrating the experimental effect at the population level. Here, we report Bayesian estimates of the population parameter with their associated uncertainty. Given 8/10 participants significant at $p = 0.01$, we can be confident that the population replication probability is greater than 49%. We would expect the majority of the population to show this result if they were tested in the same experiment.

## Supplemental information

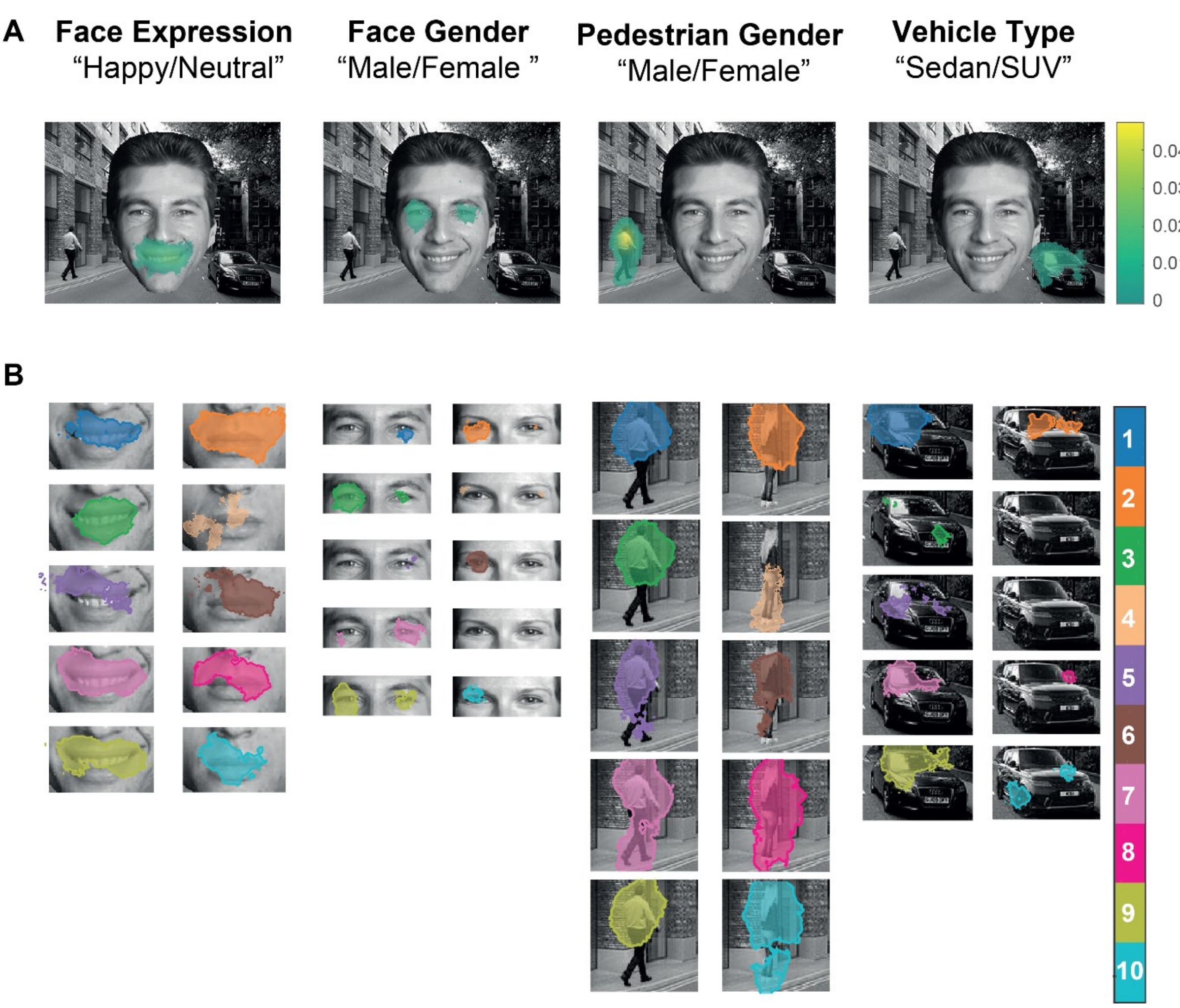

**Figure S1. Task-relevant features, related to Figure 1.** *A. Mean MI in each categorization task.* In each task (columns) and participant, for each image pixel we computed MI(<pixel visibility; correct vs. incorrect categorization>), to reveal the significant ($p < 0.05$, FWER corrected) pixels that modulate categorization accuracy. For each pixel, we computed the mean MI across all ten participants. *B. Task-relevant features in each participant.* In each task (columns) the same color-code represents the significant features for this participant. Note in each column (e.g. *pedestrian gender*) that different participants can use different (even mutually exclusive) features for the same categorization responses (e.g. for “male” vs “female pedestrian”, upper body in participants 1, 2 and 3; lower body in participant 4). From the proportion of participants who significantly used each pixel, we estimated the population prevalence, expressed as a Bayesian maximum a posteriori (MAP) [95% Highest Posterior Density Interval (HPDI)] estimate. Face expressions: MAP [95% HPDI] = 1 [0.75 - 1]. Face gender: MAP [95% HPDI] = 0.58 [0.28 – 0.83]. Pedestrian: MAP [95% HPDI] = 1 [0.75 – 1]. Vehicle: [95% HPDI] = 0.47 [0.19 – 0.75], see also *Methods, Bayesian population prevalence*.

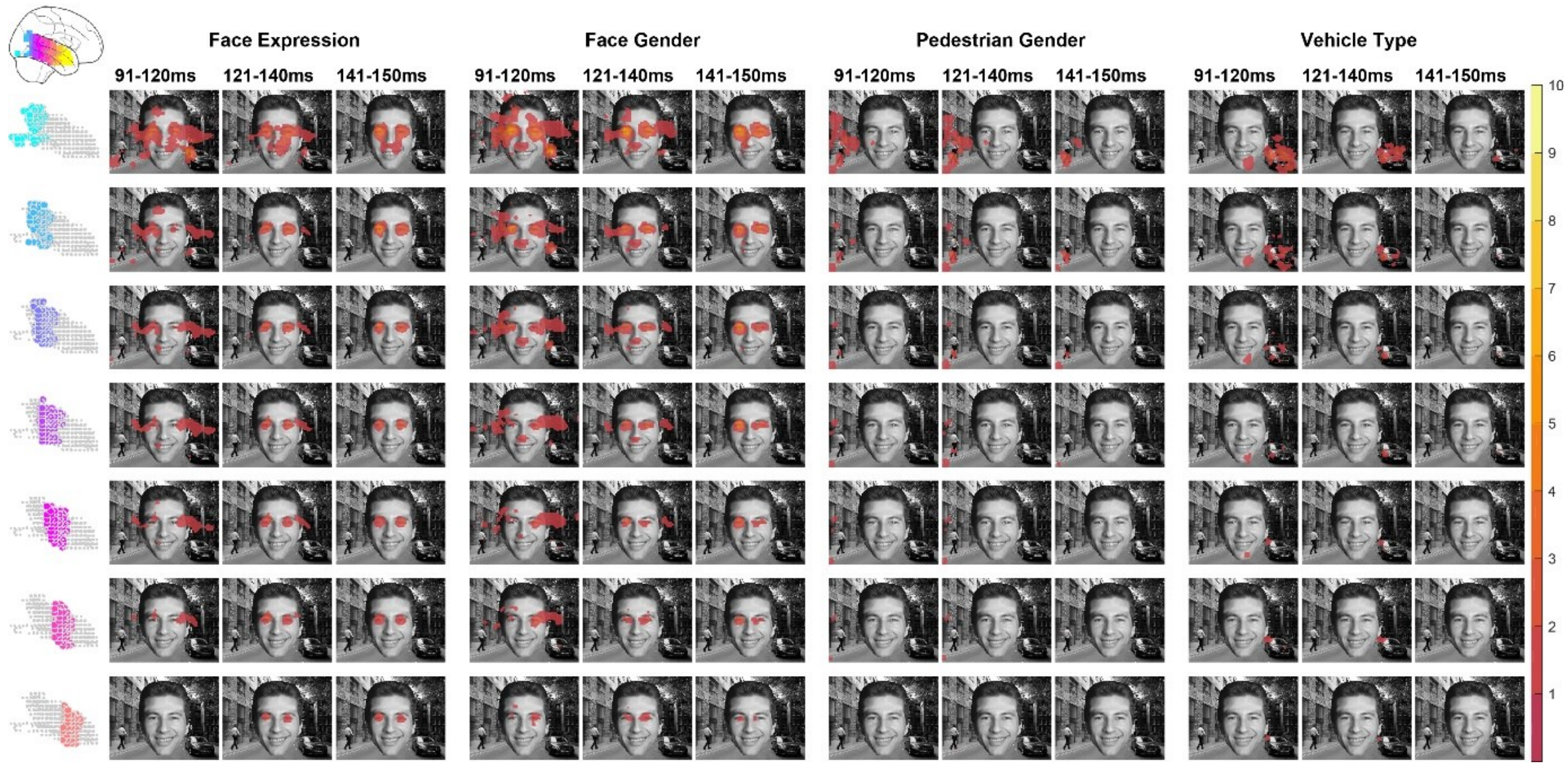

**Figure S2. Systems-level dynamic image transformations along the posterior-anterior axis of the ventral pathway, related to Figure 2.** A color-coded reference (cyan-to-yellow) for the ventral pathway sources represents their positions (back-to-front). We sliced the ventral pathway into 7 posterior-anterior clusters of increasing distances and repeated for each cluster (row), the analysis of Figure 2A to visualize the image transformations in the sources of each ventral cluster (row) and time periods (columns in each task). Images show the number of participants (color-coded red-to-yellow) that represent a given image pixel within a ventral cluster and time window and task (FDR test with q=0.05). The result show task-specific transformations of broader image representations into lower-dimensional manifolds across the clusters of the ventral pathway over ~91-150ms (i.e. in the time period when occipital cortex reduces task-irrelevant features).

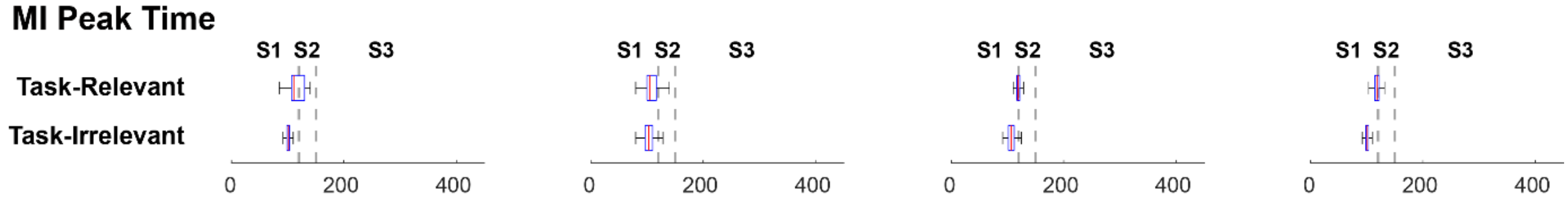

**Figure S3. Distribution of peak time of feature representation, related to Figure 3.** Across Stages 1 and 2, within 80-140 ms, for task-relevant vs. task-irrelevant features in each task.

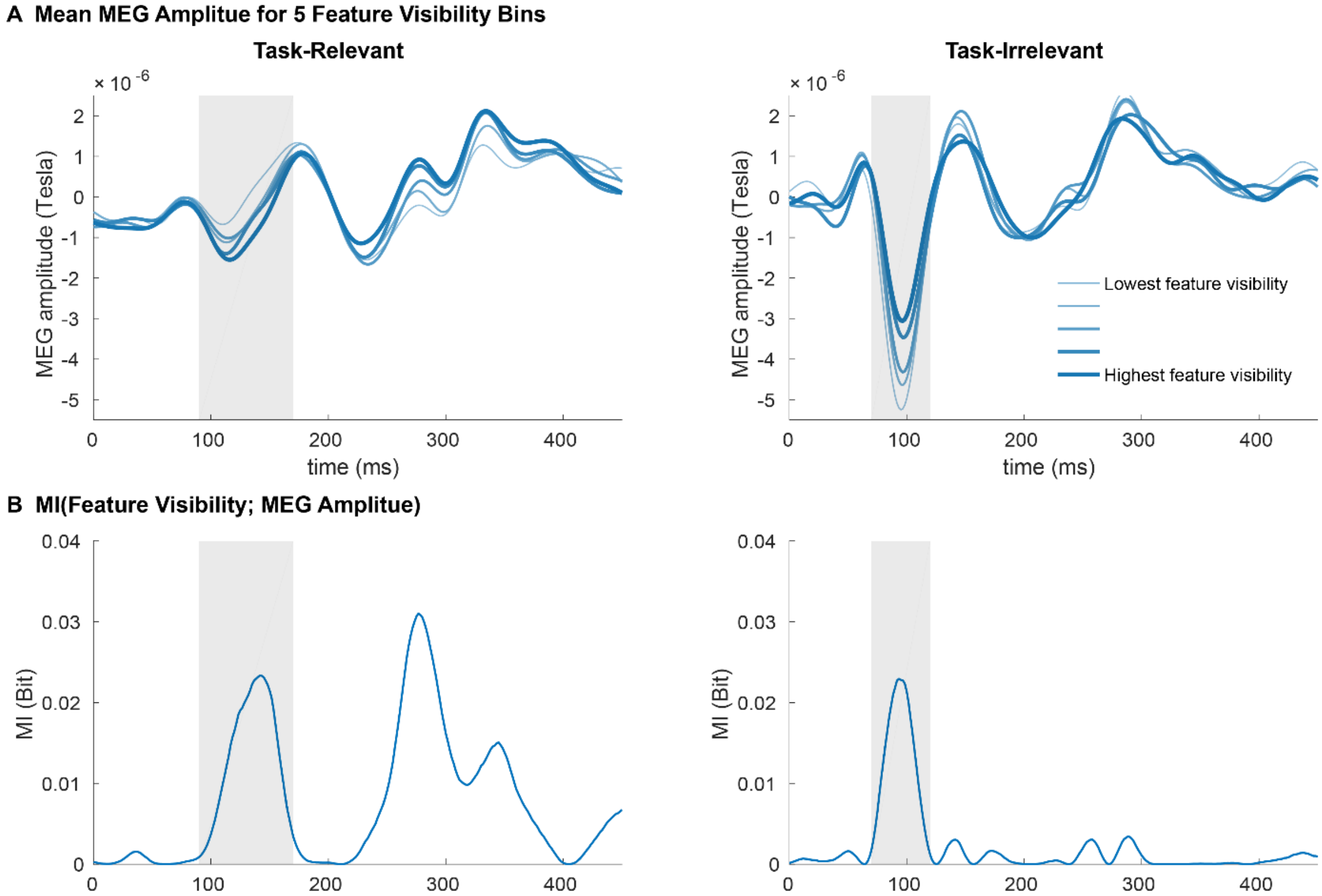

**Figure S4. Illustration of Mutual Information (MI) and opponent representation (using one example occipital source and vehicle feature relevant in vehicle-type task and irrelevant in face-expression task), related to Figure 4 and STAR Methods.** *A. Mean MEG amplitude for 5 bins of feature visibility.* We split trials into 5 bins of feature visibility. For each bin of feature visibility, we plot the corresponding mean MEG amplitude responses of an example occipital source. In the time period highlighted in grey, feature visibility correlates with MEG amplitude. However, there is an opponent representation of the feature on this occipital source depending on whether it is task-relevant vs. task-irrelevant. When the feature is task-relevant, its higher visibility corresponds to more negative MEG amplitudes. In contrast, when the feature is task-irrelevant, lower feature visibility corresponds to more negative MEG amplitudes. *B. MI(Feature Visibility; MEG Amplitude).* MI therefore quantifies the relationship between feature visibility and MEG amplitude, which is broadly called feature representation in MEG amplitude in neuroscience.

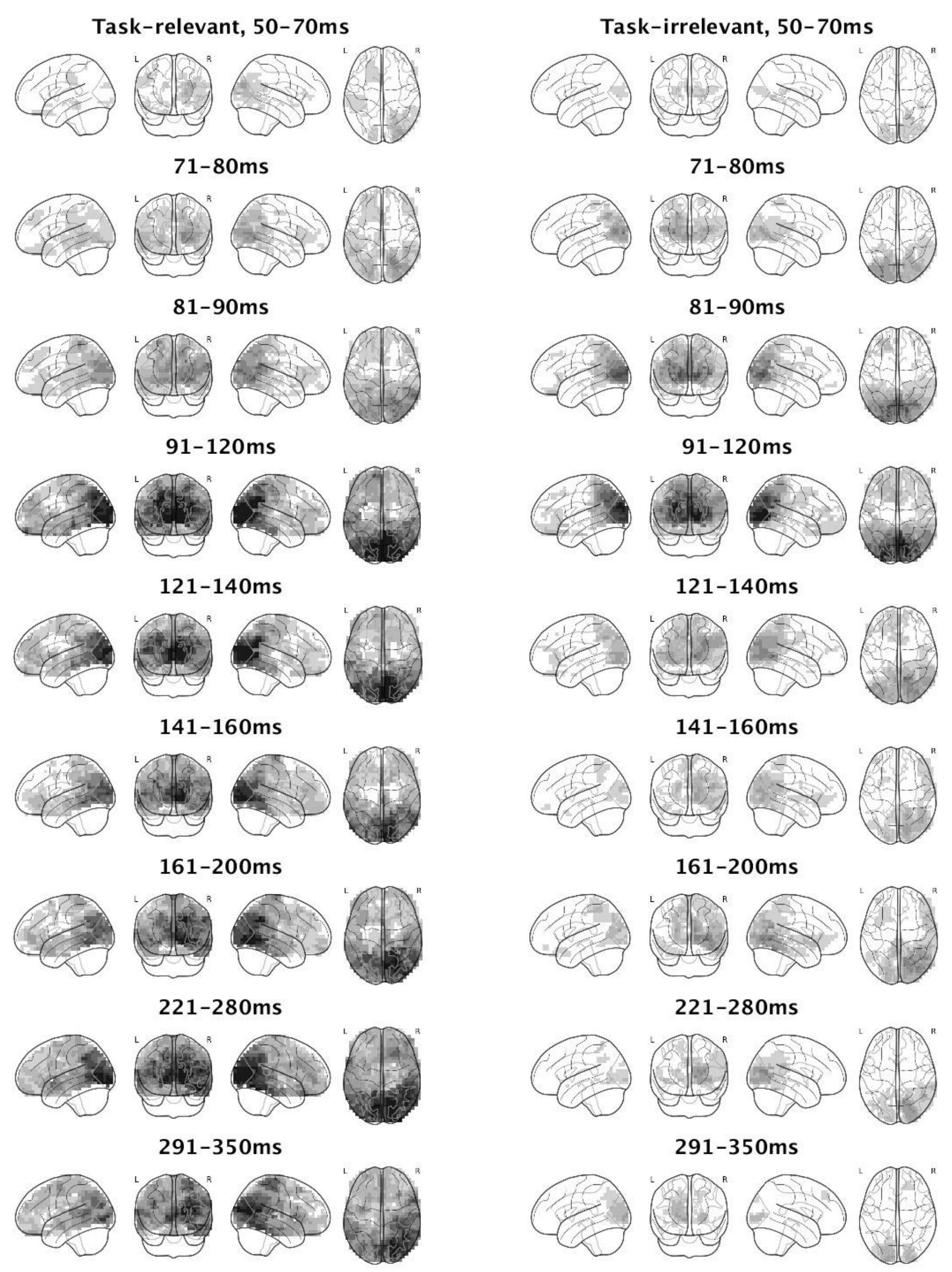

**Figure S5. Spatio-temporal dynamics of task-relevant vs. irrelevant feature representations, related to Figure 2.** This figure extends Figure 2C. Greyscale reveals the number of participants that represented at least one task-relevant vs. task-irrelevant feature in each period for each source.

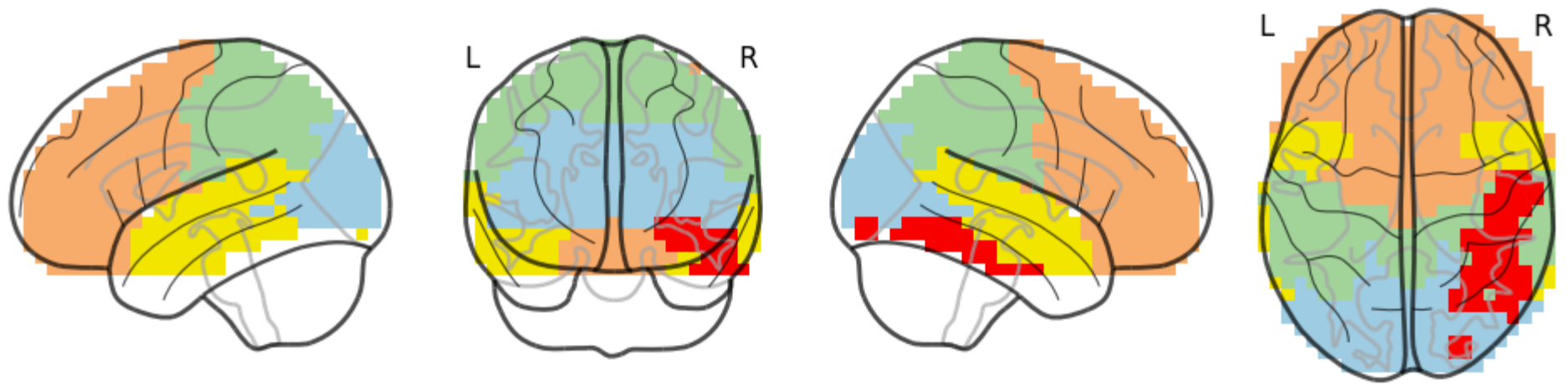

**Figure S6. The glassbrain legend, related to STAR Methods.** We plotted occipital, temporal, parietal and frontal cortices with four different colors based on the Talairach-Daemon atlas. Specially, we highlighted the right fusiform gyrus (rFG) in red.

| | Expression | Face Gender | Pedestrian Gender | Vehicle Type |
|---|---|---|---|---|
| Accuracy | 77.21% (1.45%) | 75.12% (1.92%) | 76.11% (1.56%) | 65.25% (1.91%) |
| Reaction Time | 753 ms (54.14) | 738 ms (42.24) | 779 ms (42.15) | 978 ms (50.51) |

**Table S1. Average accuracy and reaction times (standard deviation in parentheses) across participants in each categorization task, related to Results, Behavior, related to STAR Methods.**

| | Expression | Face Gender | Pedestrian Gender | Vehicle Type |
|---|---|---|---|---|
| Participant 1 | 70.31% | 68.82% | 75.20% | 69.01% |
| Participant 2 | 78.06% | 74.61% | 73.57% | 64.19% |
| Participant 3 | 83.59% | 85.74% | 80.99% | 55.17% |
| Participant 4 | 76.63% | 74.22% | 67.84% | 57.23% |
| Participant 5 | 76.30% | 74.35% | 71.74% | 60.35% |
| Participant 6 | 76.17% | 71.81% | 74.35% | 66.80% |
| Participant 7 | 84.44% | 83.07% | 83.07% | 66.54% |
| Participant 8 | 70.64% | 65.36% | 79.62% | 67.84% |
| Participant 9 | 78.26% | 75.46% | 73.31% | 72.07% |
| Participant 10 | 77.67% | 77.73% | 81.38% | 73.31% |

| | Expression (ms) | Face Gender | Pedestrian Gender | Vehicle Type |
|---|---|---|---|---|
| Participant 1 | 614 | 583 | 651 | 775 |
| Participant 2 | 1127 | 954 | 1058 | 1308 |

| Participant 3 | 654 | 690 | 730 | 846 |
|---|---|---|---|---|
| Participant 4 | 805 | 729 | 796 | 1037 |
| Participant 5 | 865 | 903 | 824 | 890 |
| Participant 6 | 684 | 816 | 745 | 1037 |
| Participant 7 | 667 | 642 | 668 | 790 |
| Participant 8 | 614 | 683 | 650 | 821 |
| Participant 9 | 596 | 554 | 725 | 928 |
| Participant 10 | 906 | 828 | 946 | 946 |

**Table S2. Per participant average accuracy and reaction times in each categorization task, related to Results, Behavior, related to STAR Methods.**